\documentclass[pra]{revtex4-1}
\usepackage{amssymb,amsmath,amsfonts,amsthm}
\usepackage{graphicx,color}
\usepackage{hyperref}
\usepackage{systeme}

\newcommand{\ket}[1]{{\left\vert {#1} \right\rangle}}
\newcommand{\bra}[1]{{\left\langle {#1} \right\vert}}
\newcommand{\braket}[2]{\langle{#1}|{#2}\rangle}

\newcommand{\beq}{\begin{equation}}
\newcommand{\eeq}{\end{equation}}
\newcommand{\bea}{\begin{eqnarray}}
\newcommand{\eea}{\end{eqnarray}}

\newcommand{\piu} {\!+\!}

\newcommand{\fig}{Fig.~}
\newcommand{\figs}{Figs.~}
\newcommand{\cf} {cf.~}
\newcommand{\ug} {\!=\!}
\newcommand{\eq}{Eq.~}
\newcommand{\eqs}{Eqs.~}

\newcommand{\rref}{Ref.~}
\newcommand{\rrefs}{Refs.~}
\newcommand{\p} {o}

\begin{document}

\author{Salvatore Lorenzo$^{(1)}$, Federico Lombardo$^{(2)}$, Francesco Ciccarello$^{(2,3)}$, G. Massimo Palma$^{(2,3)}$}

\affiliation{$^{(1)}$Quantum Technology Lab, Dipartimento di Fisica, Universit$\grave{a}$  degli Studi di Milano, 20133 Milano, Italy \& INFN, Sezione di Milano, I-20133 Milano, Italy,
$^{(2)}$Dipartimento di Fisica e Chimica, Universit$\grave{a}$  degli Studi di Palermo, via Archirafi 36, I-90123 Palermo, Italy
$^{(3)}$NEST, Istituto Nanoscienze-CNR.\\ Correspondence should be addressed to G.M.P. (e.mail: massimo.palma@unipa.it)}

\begin{abstract}

{\bf As discovered by P. W. Anderson, excitations do not propagate freely  in a disordered lattice, but, due to destructive interference, they localise. 
As a consequence, when an atom interacts with a disordered lattice, one indeed observes a non-trivial excitation exchange between atom and lattice. 
Such non-trivial
atomic dynamics will in general be characterised also by a non-trivial quantum information backflow, a clear signature of non-Markovian dynamics. 
To investigate the above scenario, we consider a quantum emitter, or atom, weakly coupled to a uniform coupled-cavity array (CCA). If initially excited, 
in the absence of disorder, the emitter undergoes  a Markovian spontaneous emission  by releasing all its excitation into the CCA (initially in its vacuum state). By introducing static disorder in the CCA 
the field normal modes become Anderson-localized, giving rise to a non-Markovian atomic dynamics. 
We show the existence of a functional relationship between a rigorous measure of quantum non-Markovianity and the CCA localization. 
We furthermore show that the average non-Markovianity of the atomic dynamics is well-described by a phenomenological model 
in which the atom is coupled, at the same time, to a single mode and to a standard - Markovian - dissipative bath.}
\end{abstract}

\pacs{03.65.Yz, 03.67.-a, 42.50.Lc}

\title{Quantum non-Markovianity induced by Anderson localization}

\maketitle

\section*{Introduction}

The dynamics of small systems interacting with {\it structured} reservoirs is
an emerging topic in the study of open quantum systems \cite{book1, book2, review1, review2, review3, review4}.
Indeed structured environments occur in several scenarios such as cavity quantum electrodynamics (QED) \cite{cqed1, cqed2,cqed3,Gao}, photonic-band-gapped materials \cite{lambro,garrawayPBG} and quantum biology \cite{quantumbio}.
While a reservoir with a flat spectral density gives rise to a {\it Markovian} Gorini-Kossakowski-Lindblad-Sudarshan (GKLS) master equation  \cite{book1, book2}, structured environments are traditionally expected to lead to non-Markovian   dynamics. This expectation relies on the common association of quantum Markovianity with a Lindblad master equation(and vice versa). Yet the notion of what is a non-Markovian open quantum dynamics has undergone a critical change in paradigm in the last few years thanks to  the introduction of a number of quantum non-Markovianity measures \cite{BreuerLP09,RivasHP10,LorenzoPP13} all based on quantum information concepts
(see also \rref\cite{review1, review2, review3, review4}). According to most of these, a Markovian dynamics does not necessarily imply a GKLS Master Equation (while the converse is true). 

A paradigmatic non-Markovian dynamics is the well-known atomic emission into a lossy cavity in the  strong coupling regime of cavity QED \cite{cqed1, cqed2, cqed3}. In this regime the atomic population exhibits damped vacuum Rabi oscillations leading to a quantum information back flow from the reservoir into the atom -- namely a distinctive trait of non-Markovian behavior \cite{BreuerLP09}. For such dynamics to occur an accurate engineering of the setup is required. In particular, in order to increase the atom-field coupling strength, it is necessary to confine the field mode within a small volume.

A conceptually different way to create a cavity-QED-like dynamics was demonstrated by Sapienza {\it et al.} in a seminal experiment \cite{lodahl}, in which a quantum emitter was coupled to a {\it disordered} photonic crystal with no cavity or resonator engineering. The mechanism responsible for such a dynamics exploits the celebrated Anderson localization \cite{anderson,Wiersma} according to which transport is inhibited in a disordered medium due to the intrinsically localized nature of all normal modes ({\it localized modes}). In this scenario a localized mode centered at the quantum emitter's site can indeed strongly couple to the emitter, much like a cavity mode does, giving rise to a cavity-QED-like dynamics. Indeed coherent polariton states have  been recently observed for a quantum dot coupled to a disordered photonic crystal \cite{Gao}.
If such an analogy holds then static disorder, whose distinctive effect is Anderson localization, should give rise to quantum non-Markovianity of the emitter dynamics. 

The above intuition is the motivation of the present paper, whose aim is to investigate such disorder-induced quantum non-Markovianity in terms of recently proposed quantum non-Markovianity measures. As a  case study we consider a quantum emitter (atom) {\it weakly} interacting with a coupled-cavity array (CCA). In the absence of disorder all the CCA modes are delocalized over the entire lattice, and -- as expected -- the atom undergoes standard Markovian spontaneous emission with the excitation propagating away  from the atom location. We introduce static disorder in the above setup in the form of random detunings of the frequencies of the array cavities distributed according to a probability distribution function (PDF) of given width. We show that the presence of disorder induces information back flow from the CCA to the atom, as witnessed by a non-monotonic time evolution of the atomic population, due to the appearance of CCA localized modes. We show that the {\em average} non-Markovianity $\bar{\mathcal N}$ grows monotonically with the disorder width, suggesting a quantitative dependence of the  non-Markovianity degree on the degree of Anderson localization. To interpret the functional form of this relationship, we introduce a simple phenomenological model in which the atom is strongly coupled to a single localized mode and, only weakly, to a dissipative reservoir. We show that this model can reproduce fairly well the functional dependence of $\bar{\mathcal N}$ on the PDF width (the latter being a measure of the disorder strength).

The present work is organized as follows. In the Section {\it The model and the open dynamics}, we define the system under investigation and the general features of the atom open dynamics. In Section {\it Localization length and non-Markovianity measures}, we illustrate the definitions of localization length and non-Markovianity measure that we adopt in out analysis. In Section {\it Non-Markovianity versus disorder strength}, we quantitatively show how the presence of disorder induces quantum non-Markovianity, deriving the functional dependance of the non-Markovianity measure on the disorder strength. In Section {\it Phenomenological model}, we introduce a simple phenomenological  model which provides a clear interpretation of the results of the previous section. We furthermore show how to determine the parameters which characterize such model in order to reproduce the ensemble averaged non-Markovianity. Finally, we discuss the findings and draw our conclusions.
\section*{results}
\begin{figure}[t!]
	\begin{center}
	\includegraphics[width=0.4\textwidth]{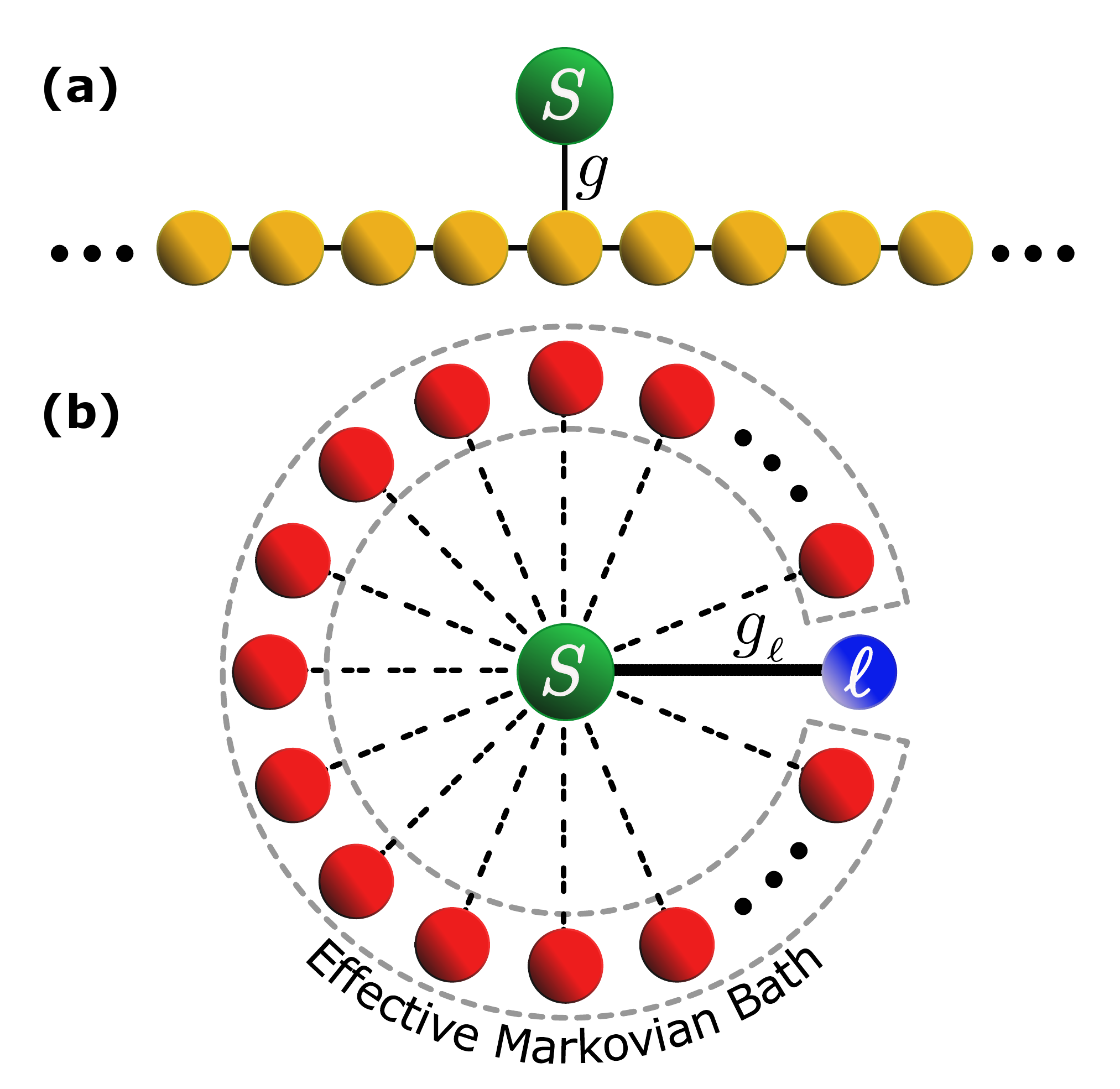}
	\caption{{\bf (a)} Sketch of the considered model: a two-level atom $S$ is coupled to the central cavity of an infinite-length CCA. {\bf (b)} Phenomenological model: $S$ is strongly coupled to a specific localized field mode $\ell$ and perturbatively to all the remaining ones. The latter field modes thus embody an effective Markovian bath.}\label{sketch}
	\end{center}
\end{figure}

\noindent{\bf {The model and the open dynamics}}\label{model1-model9}. Our model consists of quantum emitter, i.e., a two-level atom $S$, in contact with a reservoir embodied by a CCA comprising an infinite number of single-mode, coupled, lossless cavities.  A sketch of the setup is shown in \fig\ref{sketch}(a). The atomic emission process in the absence of disorder in such a system was investigated in \rref\cite{Lombardo2014}. We assume the inter-cavity coupling, i.e., the photon hopping rate $J$, to be uniform throughout the CCA, and the emitter $S$ to be coupled to the 0th cavity under the usual rotating wave approximation with coupling rate $g$. The Hamiltonian of this system reads (we set $\hbar\ug1$ throughout)
\begin{equation}
\hat H=\omega_{a} \ket{e}\bra{e}+\hat H_{f}+g\, (\hat {\sigma}_+\hat{a}_{0}+\hat {\sigma}_-\hat{a}^\dag_{0})\,,\label{H}
\end{equation}
where the free Hamiltonian of the CCA is 
\begin{eqnarray}
\hat{H}_{f}=\sum_{n=-N}^{N}\left[\varepsilon_n \hat{a}_n^\dag \hat{a}_n-J(\hat{a}_{n}\hat{a}_{n+1}^\dag+\hat{a}_{n}^\dag \hat{a}_{n+1})\right]\,,\label{Hf}
\end{eqnarray}
and where $\hat \sigma_+{=}\hat \sigma_-^\dag {=}\ket{e}\bra{g}$ are the pseudo-spin ladder operators of $S$, whose ground and excited states $|g\rangle$ and $|e\rangle$, respectively, are separated by an energy gap $\omega_a$. Here, $\hat a_n$ ($\hat a_n^\dag$) annihilates (creates) a photon in the $n$th cavity, the number of cavities of the CCA being $2N+1$ . Although \eqs(\ref{H}) and (\ref{Hf}) are written for a finite $N$, we will be ultimately interested in the limit $N{\rightarrow}\infty$ (infinite-length CCA).

We introduce {\it static disorder} in the so far uniform CCA assuming the detunings $\delta_n{=} \varepsilon_n{-}\omega_a$ [\cf\eq(\ref{Hf})]  between the nth cavity  and the atom to be random variables identically and independently distributed according to a given Probability Distribution Function (PDF). 
As mentioned in the Introduction, the presence of such disorder leads to localization of the field modes.
In the following we will analyse in detail the spontaneous emission process in this disordered environment, i.e., the irreversible dynamics that occurs when the atom is initially in its excited state $|e\rangle$ and the localised field modes are in their vacuum states $|{\rm vac}\rangle$. 
Such initial conditions,  together with the conservation of the total number of excitations, $[\hat H,\hat\sigma_{+}\hat\sigma_-\piu \sum_n\hat{a}_n^\dag\hat{a}_n]\ug0$, entail that the dynamics is restricted to the {\it single-excitation} sector of the entire Hilbert space. The effective representation of Hamiltonian (\ref{H}) in this subspace is obtained from \eq(\ref{H}) under the replacements $\hat a_n \rightarrow |n-1\rangle\langle n|$ and $\hat a_n^\dag \rightarrow |n+1\rangle\langle n|$, where $|n\rangle=\hat a_n^\dag |{\rm vac}\rangle$ is the field state with one photon in the $n$th cavity, all the remaining ones being in their vacuum state. In the following it will be convenient to rewrite the Hamiltonian in terms of normal modes. To this end, let us  consider a given noise realisation  corresponding to a fixed set of disordered cavity detunings $\{\delta_n\}$ and let $\{|\varphi_k\rangle\}$ be the eigenstates of the field Hamiltonian $\hat H_f$ such that $\hat H_f|\varphi_k\rangle{=}\omega_k|\varphi_k\rangle$, where $\omega_k$ is the normal frequency of the $k${th} mode. In term of the above states, the Hamiltonian can be rewritten in the form
\begin{equation}
\hat H{=}\omega_{a} \ket{e}\bra{e}{+}\!\!\sum_{k}\!\omega_k \ket{\varphi_k}\bra{\varphi_k}{+}\!\sum_{k}\! g_k \left(\ket{e}\bra{\varphi_k}+{\rm H.c.}\right),\label{H2}
\end{equation}
where
\begin{equation}
g_k=g\braket{0}{\varphi_k}\label{gk}
\end{equation}
is the strength of the coupling between $S$ and the $k$th field mode, $\braket{0}{\varphi_k}$ being  the probability amplitude that a photon in mode $k$ can be found in the $0${th} cavity, i.e, at  the location of the emitter $S$. With no loss of generality we will assume $g_k$ to be real.

As is well-known in standard treatments of spontaneous emission, the time-evolved atom-field state takes the form \cite{book1, book2, lambro}
\begin{equation}
\ket{\psi(t)}{=}e^{-i\hat{H}t}\ket{\psi(0)}{=}\alpha(t)\ket{e}\ket{\rm vac}{+}\sum_{k}\beta_k(t)\ket{g}\ket{\varphi_k}\label{psit}
\end{equation}
with $|\psi(0)\rangle{=}|e\rangle|{\rm vac}\rangle$. For the above state the time-dependent Schr{\"o}dinger equation yields the set of coupled differential equations
\begin{eqnarray}
i\dot{\alpha}(t){=}\omega_a \alpha(t){+}\sum_k g_k\beta_k(t),\,\,\,\,\,\,i\dot{\beta}_k(t){=}\omega_k \beta_k(t){+}g_k \alpha(t)\,.
\label{ode}
\end{eqnarray}
Integrating the latter equation one obtains a formal solution for  $\beta_k$ which is then replaced in the first equation so as to end up with an integro-differential equation for the atomic excitation amplitude
\begin{equation}
\dot{\alpha}{=}-i\omega_a \alpha(t){-}\!\!\int_0^t {\rm d}t'\sum_{k} g_k^2 e^{-i \omega_k(t{-}t')}\alpha(t')\,.
\label{alphaeq}\end{equation}
This can be solved in the Laplace space as \cite{lambro}
\begin{equation}
\tilde{\alpha}(s)=\frac{1}{s{+}i\omega_a{+}\tilde{f}(s)}\,,
\label{Lalpha}\end{equation}
where $\tilde{\alpha}(s)$ and $\tilde{f}(s)$ are the Laplace transforms of $\alpha(t)$ and $f(t)$, respectively, and where we have defined $f(t){=}\sum_k g_k^2e^{-i \omega_k t}$.

Let $\rho(t){=}{\rm Tr}_f\left\{|\psi(t)\rangle\langle \psi(t)|\right\}$ be the reduced state of $S$ at time $t$ after tracing out the field degrees of freedom. \eq(\ref{psit}) yields
\begin{equation}
{\rho}(t)=
\begin{pmatrix}|\alpha(t)|^2\rho_{ee}&\alpha(t)\rho_{eg}\\
\alpha(t)^*\rho_{ge}&(1{-}|\alpha(t)|^2)\rho_{ee}{+}\rho_{gg}\end{pmatrix}\,,
\label{TLSmap}\end{equation}
where $\rho_{jk} =\langle j|\rho(0)|k\rangle$ with $j,k=g,e$ the entries of the initial atomic density matrix $\rho(0)$. The dynamics of the emitter is therefore
an amplitude-damping channel \cite{NielsenCbook}, as entailed by the form of Hamiltonian (\ref{H2}).

In the absence of disorder, i.e., when $\delta_n{=}\delta$ for all $n$, the free Hamiltonian of the CCA (\ref{Hf}) reduces to a standard uniform tight-binding model. 
This can be exactly diagonalised and the resulting single-excitation eigenstates are plane waves (Bloch functions) given by $\ket{\varphi_k}{=}1/\sqrt{2N{+}1}\sum_n e^{ikn}\ket{n}$ with associated eigenvalues $\omega_k{=}2J\cos k$, where $k{=}2\pi m/(2N{+}1)$ and $m$ is an integer index (since we are interested in the emission process into an infinite-length CCA, we conveniently assume {\it cyclic} boundary conditions for the CCA throughout). In this case, due to the cosine form of the dispersion law with its ensuing finite band, the atom  emission is in general non-monotonic in time and can even be fractional, i.e., part of the emitter initial excitation may not be released into the CCA \cite{Lombardo2014}. However, in the {\it weak-coupling} regime  $(g{\ll}J)$ and for $\delta{=}0$, i.e., for $\varepsilon_n{=}\omega_a$ [\cf\eqs(\ref{H}) and (\ref{Hf})], the CCA's dispersion is approximatively linear and with an infinite energy band. In this regime, $\alpha(t)=e^{-\frac{g^2}{2J}t}$ \cite{Lombardo2014,ciccarello2016}, namely the atom undergoes standard exponential decay, its open dynamics being thus fully Markovian. In the following, our main goal is to study how this weak-coupling regime is affected by the introduction of disorder into the CCA. In particular, we will analize the non-Markovianity induced by the disorder.
\\
\\
\noindent{\bf Localization lenght and non-Markovianity measure}. Let us now define the two key quantities that will play a central role in our analysis, namely the localization length and the non-Markovianity measure.
\\
\noindent {\em Localization length.}-  \label{ANDERSON} Localization is a well-known effect of static disorder in a one-dimensional lattice. In the setup considered in this paper, it originates from the random distribution of cavity detunings $\delta_n$, which makes  {\it each} CCA eigenstate $\ket{\varphi_k}{=}\sum_n c_{kn}\ket{n}$  exponentially localized around a lattice site, i.e., for each $k$ there exists a site $x_0$ such that $\langle x|\varphi_k\rangle\sim e^{-\frac{|x-x_0|}{\lambda_k}}$ \cite{Ishii1973,Thouless1974}. This occurs no matter how weak the disorder is. The characteristic length of such exponential decay
for the $k$th eigenstate is called localization length $\lambda_k$ and can be defined in different ways \cite{Molinari1993,Thouless1972,Kramer1993}. 
A commonly used definition in terms of generalized entropies is \cite{Molinari1993}
\begin{equation}
\lambda_k^{(q)}{=}\left(\sum_n |c_{kn}|^{2q}\right)^{\frac{1}{1{-}q}}\,,\label{lambdaq}
\end{equation}
which for $q = 1$ and $q=2$ reduces to the so called information length and participation ratio, respectively.

An alternative definition \cite{Thouless1972,Herbert1971,Ishii1973} 
is expressed in terms of the residues of the Green function (resolvent) associated with the lattice Hamiltonian, i.e., $\hat H_f$ in our case. It reads
\begin{equation}
\tilde\lambda_k=\frac{2N}{ \sum_{j \neq k} \log|\omega_k-\omega_j|}\,.
\label{lambdaTH}\end{equation}
In this work we will quantify the localization length in terms of  $\lambda_k^{(2)}$ and $\tilde\lambda_k$.
\\
\noindent{\em Non-Markovianity measure}.-
\label{NM}
To quantify the amount of quantum non-Markovianity of an open dynamics a number of theoretical measures have been put forward in the last few years  \cite{review1, review2, review3, review4, BreuerLP09, RivasHP10,LorenzoPP13}. Such measures have been used to identify the regions in the parameters space corresponding to Markovian and non-Markovian  dynamics for a variety of environmental models \cite{model1, model2, model3, model4, model5, model6, model7, model8, model9}. Different measures lead in general to non-equivalent partitions. However the amplitude damping channel -- which is the class of open dynamics involved in our case [\cf\eq(\ref{TLSmap})] -- is a relatively simple one since for this channel the non-Markovianity measures introduced in \rrefs\cite{BreuerLP09, RivasHP10,LorenzoPP13} all lead to the same criterion for the occurrence of non-Markovian behaviour: non-Markovianity occurs if and only if the time derivative of $|\alpha(t)|$ [\cf\eqs(\ref{psit}) and (\ref{TLSmap})] is positive at some time. In other words, the dynamics is non-Markovian iff the atomic excited-state population grows at some time (in contrast to the Markovian case where it monotonically decreases with time). In particular, according to the measure in \rref\cite{BreuerLP09}, this increase of atomic excitation corresponds to the occurrence of information back flow from the reservoir to the open system. 

Throughout this work, mainly for its computational convenience, we will adopt the non-Markovianity measure introduced in \rref\cite{LorenzoPP13}. This is formulated in terms of the time evolution of the volume of accessible states of the open system $V(t)$. As this volume can only decrease with time for a Markovian dynamics (in particular one governed by the GKLS master equation) the amount of quantum non-Markovianity is measured by \cite{LorenzoPP13}
\begin{equation}
\mathcal{N}_V=\frac{1}{V(0)}\int_{\partial_t{V(t)}>0}\!{\rm d}t\,\,\partial_t{V(t)}\,,
\label{NM_nostra}
\end{equation}
where the integral is over the time domains in which $V(t)$ increases (as indicated by the subscript $\partial_t{V(t)}>0$). For a dynamical map of the form \eqref{TLSmap}, $\mathcal{N}_V$ is explicitly given by \cite{LorenzoPP13}
\begin{equation}
\mathcal{N}_V=\int_{\partial_t{|\alpha(t)|}>0}\!{\rm d}t\,\frac{\rm{d}|\alpha(t)|^4}{\rm{d}t}\,.
\label{NMV}
\end{equation}
Clearly $\mathcal{N}_V>0$ if and only if $|\alpha(t)|$  grows at some times. If instead $|\alpha(t)|$ monotonically decreases with time then $\mathcal{N}_V=0$ corresponding to a Markovian behavior.
The measure so defined diverges if $|\alpha(t)|$ exhibits stationary oscillations, e.g. in the case of an atom undergoing vacuum Rabi oscillations due to its strong coupling with a lossless cavity mode. To get around this drawback, which can occur in our setup since, as we will show, the atom can strongly couple to a localized mode, we 
rescale the non-Markovianity measure as 
\begin{equation}
\mathcal{N}\;{=}\frac{\mathcal N_V}{\left|\int_{\partial_t{|\alpha(t)|}<0}\frac{\rm{d}|\alpha(t)|^4}{\rm{d}t}\right|}\,,
\label{generalRNM}
\end{equation} 
where the denominator is the analogue of \eq(\ref{NMV}) but now evaluated with respect to time {\it decreases} of $V(t)$. For a Markovian dynamics, $|\alpha(t)|$ monotonically decreases with time, hence $\mathcal N=0$, since the numerator in \eq(\ref{generalRNM}) vanishes while the denominator diverges. In contrast, for a Jaynes-Cummings dynamics -- which can be regarded as an extreme instance of non-Markovianity -- $|\alpha(t)|$ undergoes undamped Rabi oscillations. In similar cases the numerator equals the denominator yielding $\mathcal N=1$. Similar ways of rescaling non-Markovianity measures to avoid divergences have been used in the literature \cite{RivasHP10}.
\\
\\
\noindent
{\bf Non-Markovianity versus disorder}.
In this section we will link the average non-Markovianity of our disordered model with the amount of disorder. From now on, we  assume the detunings $\delta_n\ug\omega_a{-}\varepsilon_n$ to be independent random variables distributed according to a PDF  $p(\delta)$, independent of the cavity site $n$. To explore how our results depend on specific probability distributions we  consider a Gaussian distribution $p_g(\delta)$ as well as a Cauchy distribution $p_c(\delta)$ both centered at $\delta{=}0$ (corresponding to $\varepsilon_n{=}\omega_a$) and defined by
\begin{equation}
p_{g}(\delta){=}\frac{e^{-\frac{\delta^2}{2 \sigma ^2}}}{\sqrt{2 \pi } \sigma }\;,\;\;\;\;\;
p_{c}(\delta){=}\frac{\Gamma }{\pi  \left(\Gamma ^2{+}\delta ^2\right)}\,.
\label{PDFs}\end{equation}
To be consistent in our comparison  we constrain the Cauchy distribution width $\Gamma$ to fulfill $\int_{-2\sigma}^{2\sigma}p_{c}(\delta){\rm d}\delta{=}\int_{-2\sigma}^{2\sigma}p_{g}(\delta){\rm d}\delta$, meaning that the probability that $-2\sigma{\le}\delta{\le}2\sigma$ is the same with either PDF. This constraint yields $\Gamma{=}2 \sigma \cot[\pi/2 {\rm Erf}(\sqrt{2})]{\simeq}0.143\sigma$, a condition that will be fulfilled throughout.

It is important to stress that since the $\delta_n$ are {\em time-independent} random variables, each particular pattern of detunings leads to a different time dependance of the atomic spontaneous emission and therefore to a different amount of non-Markovianity. We will thereby investigate how the {\em ensemble averaged} non - Markovianity depends on the amount of disorder. We evaluate numerically the ensemble averaged non-Markovianity, for a given $N$ and disorder strength $\sigma$, 
by generating a set of detunings $\{\delta_n\}$ according to the chosen PDF of width $\sigma$ and compute $\mathcal N$ through \eq\eqref{generalRNM}. We then iterate this procedure, with a different set of detunings distributed with the same PDF, for a sufficiently large number of times (typically of the order of
thousands) and eventually evaluate the ensemble-averaged non-Markovianity measure $\overline {\mathcal N}$. In each single realization, we track the dynamics up to time  $t{=}T$, where $T$ is the time at which the atom would release 99\% of its initial excitation in the absence of disorder (i.e., for $\sigma{=}0$). This implies that the chosen CCA size must be at least equal to $\upsilon_{\rm max}T$, where $\upsilon_{\rm max}{=}2J$ is the maximum photon group velocity. Below such size unwanted boundary reflections due to the CCA finiteness would occur. In the presence of noise, we require the array size $N$ to exceed this threshold, checking out that $N$ and $T$ are such that the leftmost and rightmost cavities never get excited during the atom emission.
Throughout, we measure energies in units of $J$ (inter-cavity coupling rate) and set $g{=}0.1$ to guarantee weak-coupling conditions \cite{Lombardo2014}.
\begin{figure}[h!]
	\begin{center}
	\includegraphics[width=0.4\textwidth]{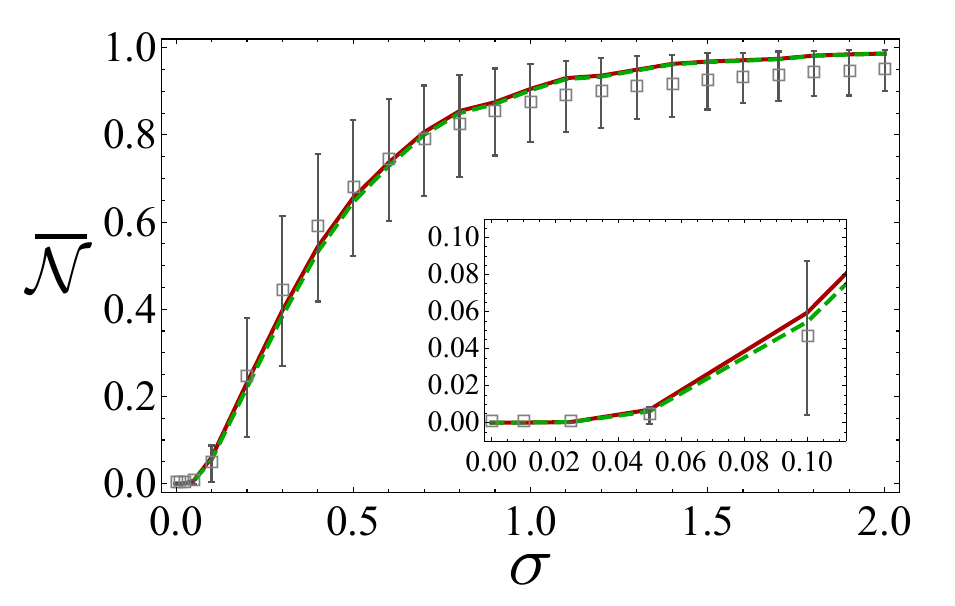}
	\caption{Ensemble-averaged non-Markovianity measure $\overline{\mathcal N}$ versus width $\sigma$ for a gaussian PDF resulting from numerical simulations of the full model (grey points) and the phenomenological model (curves). We set $N{=}1000$. For each value of $\sigma$, averages were performed over $4{\times}10^3$ different realizations of disorder. The numerical points (in grey) are shown with the associated error bars (calculated as the mean absolute deviations). The curves correspond to the outcomes of the phenomenological model with the atom-localized-mode coupling strength calculated as $g_\ell{=}2g/\sqrt{\lambda_\ell^{(2)}}$ (red-dotted) and  $g_\ell{=}1.5g/\sqrt{\tilde{\lambda}_\ell}$ (green-dotdashed). Inset: behavior of $\overline{\mathcal N}$ for low values of $\sigma$.}
	\label{NMgauss}
	\end{center}
\end{figure}

\begin{figure}[h!]
	\begin{center}
	\includegraphics[width=0.4\textwidth]{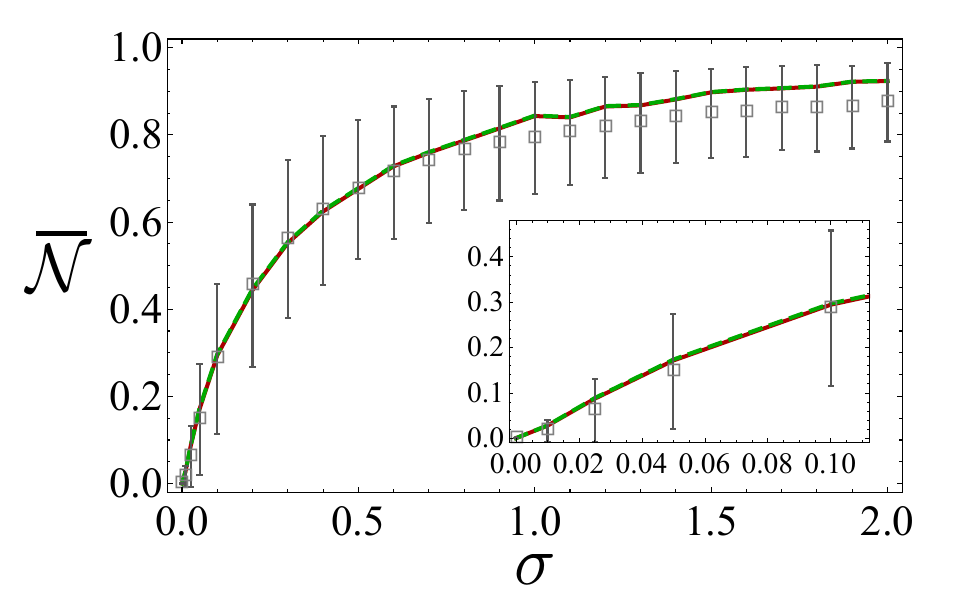}
	\caption{Ensemble-averaged non-Markovianity measure $\overline{\mathcal N}$ versus width $\sigma$ for a Cauchy PDF resulting from numerical simulations of the full model (grey points) and the phenomenological model (curves). We set $N{=}1000$. For each value of $\sigma$, averages were performed over $4{\times}10^3$ different realizations of disorder. The numerical points (in grey) are shown with the associated error bars (calculated as the mean absolute deviations). The curves correspond to the outcomes of the phenomenological model with the atom-localized-mode coupling strength calculated as $g_\ell{=}2g/\sqrt{\lambda_\ell^{(2)}}$ (red-dotted) and  $g_\ell{=}1.5g/\sqrt{\tilde{\lambda}_\ell}$ (green-dotdashed). Inset: $\overline{\mathcal N}$ for low values of $\sigma$.}
	\label{NMcauchy}
	\end{center}
\end{figure}

In \fig\ref{NMgauss} we plot the ensemble-averaged non-Markovianity measure $\overline {\mathcal N}$ versus the disorder strength $\sigma$ for a Gaussian PDF. As expected, f\p r $\sigma=0$ the behavior is Markovian since we retrieve a uniform CCA weakly coupled to the atom, while, in the presence of disorder, non-Markovianity always occurs for any finite value of $\sigma$. In particular, $\overline {\mathcal N}$ monotonically increases with $\sigma$ and eventually saturates to $\overline {\mathcal N}{=1}$ for very large $\sigma$ (we recall that $\sigma$ is expressed in units of $J$). Such asymptotic behaviour can be easily understood as follows: for very strong disorder each CCA normal mode amplitude is non zero only at a single lattice site. In this regime thereby the atom couples only to the field mode localised at its site, which results in the pure vacuum Rabi oscillations of a Jaynes-Cummings dynamics yielding $\overline {\mathcal N}{=1}$ [see Subsection {\it Non-Markovianity measure} and \eq(\ref{generalRNM})]. 
The above features persist if, instead of a Gaussian distribution, we assume a Cauchy PDF as shown in \fig \ref{NMcauchy}. Here again $\overline {\mathcal N}$ exhibits a monotonic increase with $\sigma$ saturating to $\overline {\mathcal N}=1$ for large $\sigma$. The dependance of $\overline {\mathcal N}$ on $\sigma$ is very similar to the Gaussian case (\cf\fig\ref{NMgauss}) apart from a somewhat steeper increase for small $\sigma$ and a somewhat slower convergence to $\overline {\mathcal N}{=1}$.

Static disorder thereby induces non-Markovianity  and we have  rigorously proved the intuitive expectation that photon localisation induces an information back flow from the reservoir to the system. While the limits of vanishing and very large disorder strength are clear, less obvious is interpreting the shape of function $\overline {\mathcal N}(\sigma)$ in \figs\ref{NMgauss} and \ref{NMcauchy}. In the next section, we thus formulate a phenomenological model that reproduces to a large extent the relationship $\overline {\mathcal N}(\sigma)$.
\\
\\
\noindent{\bf Phenomenological model.} 
\label{effective}
We now introduce an effective phenomenological model able to reproduce the behaviour of the average non-Markovianity measure [\cf \figs \ref{NMgauss} and \ref{NMcauchy}].
The basic idea behind our model is that (a) the atom interacts strongly and coherently with a single normal mode of the field but weakly and dissipatively with all the remaining ones, which we treat as a markovian environment (b) the former (latter) interaction is the dominant one for large (small) values of $\sigma$ while a competition between the two take place in the intermediate regime. As we are going to show, a noteworthy feature of our model is that its key parameters depend just on the variance of the disorder of the array as such. 

To define our model, let us reconsider \eqs \eqref{ode}. To derive \eq(\ref{alphaeq}) we solved for all the $\beta_k(t)$'s in the second identities of \eqs\eqref{ode} to substitute them  in the differential equation for $\alpha(t)$. Let us now instead solve for all the $\beta_k(t)$'s {\em but one}, labelled $k{=}\ell$.  In the interaction picture, then,  the following pair of equations holds
\begin{eqnarray}
\dot{\alpha}(t)&{=}&{-}ig_\ell e^{-i(\omega_\ell{-}\omega_a) t}\beta_\ell(t) {-}\sum\limits_{k\neq \ell} g_k^2\!\!\int_0^t\!\! {\rm d}t' e^{-i(\omega_k{-}\omega_a)(t{-}t')} \alpha(t')\,,\label{alpha2}\\
\dot{\beta}_\ell(t)&{=}&{-}ig_\ell e^{-i (\omega_a-\omega_\ell)t}\alpha(t)\,.
\label{beta2}
\end{eqnarray}
Let us next assume that the atom is strongly coupled to the $\ell$th mode but only weakly to modes $k{\neq}\ell$. These latter modes can then be reasonably treated as a Markovian bath as pictorially sketched in \fig\ref{sketch}(b).
Accordingly, in the spirit of the Markov or Weisskopf-Wigner approximation, we set the upper integral limit to infinity in \eq(\ref{alpha2}) 
and replace $\alpha(t'){=}\alpha(t)$. Such integral then becomes
\begin{equation}
\int_0^\infty\!\! {\rm d}\tau e^{-i(\omega_k{-}\omega_a)\tau}=\pi \delta(\omega_k{-}\omega_a)-i \mathcal{P}\left(\frac{1}{\omega_k{-}\omega_a}\right)\,,
\end{equation}
where the Cauchy principal part term, leads to a frequency shift that, for our purposes, can be neglected. Defining the relaxation rate as
\begin{equation}
\gamma=\pi \sum\limits_{k\neq \ell} g_k^2 \delta(\omega_k{-}\omega_a)\label{rate}\,,
\end{equation}
\eq(\ref{alpha2}) becomes $\dot{\alpha}(t){=}{-}ig_\ell e^{-i(\omega_\ell{-}\omega_a) t}\beta_\ell(t){-} \gamma\alpha(t)$ which, together with \eq(\ref{beta2}), now form a closed system of equations in $\{\alpha(t),\beta_\ell(t)\}$. These are equivalent to the master equation (in the Schr{\"o}dinger picture)
\begin{eqnarray}
\dot{\rho}_{S\ell}&=&-i \left[\omega_a |e\rangle\langle e|{+}{\omega_\ell} |\varphi_\ell\rangle\langle\varphi_\ell|+
	g_\ell(\hat \sigma_+\hat  a_\ell{+}{\rm H.c.}),\rho_{S\ell}\right] {+}\gamma\left(\hat\sigma_-\rho_{S\ell}\hat\sigma_+ 
	-\frac{1}{2}\lbrace\hat\sigma_+\hat\sigma_-,\rho_{S\ell}\rbrace\right)\,,\label{ME}
\end{eqnarray}
where $\rho_{S\ell}$  the joint state of the emitter $S$ and mode $\ell$ (curly brackets stand for the anticommutator, while $\hat a_\ell$ is the annihilation operator of the localized mode $\ell$).

Note that a major difference with the usual dissipative Jaynes-Cummings model, where the emitter is coupled to a lossy cavity mode, is the fact that here
the dissipator [second term of \eq(\ref{ME})] acts on the atom $S$. After straightforward calculations, one finds that the reduced 
state of $S$ following from master equation (\ref{ME}) evolves according with an amplitude channel \eq\eqref{TLSmap} with the excitation amplitude given by
{
\begin{equation}
\alpha(t)=e^{-\frac{1}{4}t(\gamma\!-\!2i(\omega_\ell\!+\!\omega_a))}\left[\cos\left(\tfrac{\Omega t}{4}\right)-\tfrac{\mu}{\Omega}\sin\left(\tfrac{\Omega t}{4}\right)\right]
\label{alphaEFF}\end{equation}
}
with $\mu=\gamma-2i(\omega_\ell\!-\!\omega_a)$ and $\Omega=\sqrt{16 g_\ell^2-\mu^2}$.
\\
\\
\noindent {\em Non-Markovianity.} Since, for the above phenomenological model, the emitters dynamics is described by an amplitude damping, the amount of non-Markovianity can be conveniently quantified in terms of the measure introduced in \rref\cite{LorenzoPP13} and described in the previous section. From \eqs(\ref{NMV}) and (\ref{generalRNM}) one finds (see Appendix for details)
\begin{equation}
\mathcal N=\frac{\mathcal N_V}{\mathcal N_V+1}\,.\label{NV+1}
\end{equation}
When the emitter is resonant with the effective cavity mode, i.e., for $\omega_\ell=\omega_a$, ${\mathcal N_V}$  can be calculated analytically (see Appendix) and one obtains

\begin{figure}[htbp]
	\begin{center}
	\includegraphics[width=0.4\textwidth]{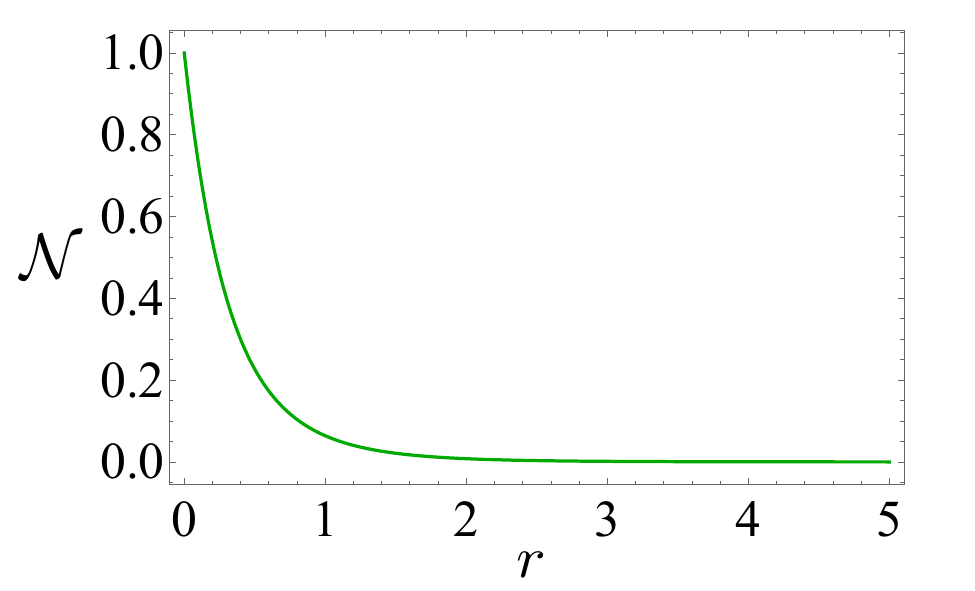}
	\caption{Non-Markovianity measure $\mathcal N$ of the phenomenological effective model as a function of $r=\gamma/g_\ell$ according to \eqs(\ref{NV+1}) and (\ref{NMmodel}).}\label{figNMmodel}
	\end{center}
\end{figure}

{\begin{equation}
\mathcal{N}_V{=}e^{-4r t_0}\times
\left\{
\begin{array}{lr}
        \left(e^{\tfrac{4 \pi  r}{\Delta }}{-}1\right)^{-1}& \text{for }\,\,0{\le} r{<}2\sqrt{2}\\
        \\
        \left(1{-}e^{-\tfrac{4 \pi  r}{\Delta }}\right)^{-1}&\text{for }\,\, 2\sqrt{2}{\le} r{<}4\\
        \\
        1 \hspace{1cm} \text{for } \,\,r{\ge} 4
\end{array}
\right .\label{NMmodel}
\end{equation}}
with $r{=}\gamma/g_\ell\;,\;\;\Delta{=}\sqrt{16-r^2}\;\text{and}\;\;t_0{=}{4}/{\Delta }\tan^{{-}1}\!\left(\tfrac{2 \Delta  r}{r^2{-}\Delta ^2}\right)$. As shown in \fig\ref{figNMmodel}, $\mathcal N$ monotonically decreases with the ratio $\gamma/g_\ell$. For $g_\ell\gg \gamma$ the non-Markovianity ${\mathcal N}\rightarrow 1$ since the effective model tends to the lossless Jaynes-Cummings model, while for $g_\ell\ll \gamma$,  $\mathcal N\simeq 0$, since in this regime the coupling to the Markovian bath dominates.

A remarkable feature of our model is the fact  that $\mathcal N$ is {\it always} non-zero except for $g_\ell=0$ (i.e., for an infinite value of $r=\gamma/g_\ell$). Such feature does not occur in other models of non-Markovian dynamics \cite{models} -- in particular it does not occur for the dissipative Jaynes-Cummings model -- where instead a finite threshold separates the Markovian from the non-Markovian regime. Observe that the absence of such a threshold for our phenomenological model is in agreement with the behavior of the ensemble-averaged non-Markovianity $\overline {\mathcal N}(\sigma)$ of \figs\ref{NMgauss} and \ref{NMcauchy} (as pointed out earlier this is is always non-zero whenever disorder is present).
\\
\\
\noindent {\em Parameters of the effective model.}
We now discuss a key point of our model, namely the the link between its parameters and the disordered of the system whose ensemble averaged non-Markovianity we want to reproduce. The only two parameters which enter the master equation of the effective model are the localised mode-emitter coupling $g_\ell$ and the decay rate $\gamma$. We will show how they both depend  on $\sigma$ only.
By definition, we identify the local mode $\ell$ as the normal mode of the free disordered CCA -- i.e., for a {\em specific} realization of disorder -- that maximizes $|\braket{\varphi_k}{0}|/{|\omega_k{-}\omega_a|}$, namely the absolute value of the ratio between the probability amplitude  at the atomic location $x=0$ and the detuning from the emitter's frequency. 
This criterion takes into account the tradeoff between the fact that a mode matching the atom frequency but localized around a site far from $x=0$ will be weakly interacting with $S$ while field modes strongly overlapping the 0th site could be far detuned, leading again to weak coupling. 
On a more formal ground, this definition relies on the first-order perturbation theory (taking the atom-CCA interaction as the interaction Hamiltonian). According to this, the correction of an eigenstate of the unperturbed Hamiltonian is indeed $\propto |\braket{\varphi_k}{0}|/{|\omega_k{-}\omega_a|}$.
Once the localised mode is identified we assume  the coupling strength $g_\ell$ [see \eq(\ref{ME})], to be inversely proportional to the square root of the localization length associated with the $\ell$th mode, i.e.,
\begin{equation}
g_\ell=C\frac{g}{\sqrt{\lambda_\ell}}\,,
\label{gell}\end{equation}
where $\lambda_\ell$ can be quantified through \eq(\ref{lambdaq}) or \eq(\ref{lambdaTH}) while the {\it$\sigma$-independent} constant $C$ has been determined in order to optimize the fitting.
The reason of this choice is that $g_\ell$ behaves in this way like the coupling strength between an atom and a cavity mode, which is well-known to be inversely proportional to the square root of the effective cavity volume. Here, the localisation length behaves as an effective cavity volume, which is a description that found experimental confirmation \cite{lodahl}. The constant $C$ will in general depend on the adopted definition of localisation length.

To compute the rate $\gamma$ [\cf\eq(\ref{rate})], we replace \eq(\ref{gk}) into \eq(\ref{rate}), integrate over $\omega_k$ and end up with
$$\gamma=\pi g^2 \left[ |\braket{0}{\varphi_k}|^2|_{\omega_k=\omega_a} \right] \rho(\omega_a)\,,$$
where $\rho(\omega)$ is the density of states of the disordered CCA in the considered realization of noise. In practice, the numerical calculation of $\gamma$ is carried out by selecting all the modes $k$, but $\ell$, whose energies lie within the interval $\omega_a-g \le \omega_k\le \omega_a+g$. Accordingly  $\rho(\omega_a)$ is computed by dividing the total number of such modes by 2$g$.

To obtain the average non-Markovianity for a given value of disorder strength $\sigma$, we take the ensemble averages of $g_\ell$ and $\gamma$ (whose calculation for a specific noise realization has been described above) and replace them into \eq(\ref{NMmodel}). Recalling that \eq(\ref{NMmodel}) holds for $\omega_\ell=\omega_a$, note that this procedure implies enforcing that the local mode $\ell$ be perfectly resonant wit the atom, i.e., $\omega_\ell=\omega_a$. Despite $\omega_\ell$ grows with $\sigma$, we indeed numerically checked that in the range $0\le\sigma\le 2$ this assumption is effective since the average $|\omega_\ell-\omega_a|$ is in the worst case only a few tenths of $\sigma$. In \figs\ref{NMgauss} and \ref{NMcauchy}, we compare the behavior of the non-Markovianity measure predicted by the phenomenological model with that of the full model in the case of a Gaussian and a Cauchy PDF, respectively. In either case, we calculated $g_\ell$ on the basis of \eq(\ref{gell}) using both $\lambda_\ell^{(2)}$ and $\tilde\lambda_\ell$ [\cf\eqs(\ref{lambdaq}) and (\ref{lambdaTH})] to define the localization length [each with a suitable $C$ factor, see \eq(\ref{gell})]. In each case, the phenomenological model predictions are clearly in good agreement with those of the full model. In passing, we mention that the agreement can be further (although slightly) improved if the constraint $\omega_\ell=\omega_a$ is relaxed. In such a case, however, the non-Markovianity measure of the effective model can no longer be calculated analytically as in \eq(\ref{NMmodel}).

\section*{\bf Discussion}

In this paper, we investigated the open dynamics of a quantum emitter in dissipative contact with a disordered environment, here embodied by a CCA,  in the weak-coupling regime. In absence of disorder, the atom undergoes standard, hence fully Markovian, spontaneous emission. When disorder is present in the form of random cavity detunings, the CCA exhibits Anderson localization since all of its normal modes are localized. By using a rigorous non-Markovianity measure, we showed that such photon localization induces non-Markovianity of the atom's emission. Intuitively, this arises from light localization, which enables information back flow from the photonic reservoir to the emitter. We found that non-Markovianity takes place for any finite disorder strength, no matter how small it is, in contrast to other environmental models where instead non-Markovian behavior occurs only beyond a finite threshold. The ensemble-averaged non-Markovianity measure $\bar{\mathcal N}$ grows with the disorder width $\sigma$ until it saturates to a value corresponding to vacuum Rabi oscillations since, for large disorder, the atom is coupled only to a single localized normal mode of the CCA. In order to understand the functional dependance of $\bar{\mathcal N}$ versus $\sigma$, we formulated a phenomenological effective model where the atom is coherently coupled to a single mode and weakly to a Markovian bath. Once the dependence of its parameters  on the disorder strength $\sigma$ are heuristically defined, this model was shown to predict the behaviour of the average non-Markovianity of our original disordered system.

\acknowledgements
 This work is supported by the EU Collaborative Project TherMiQ (Grant Agreement 618074).
\bigskip

\section*{Additional informations}

\noindent{\bf Authors contribution:}F.C and G.M.P. conceived the problem, guided the the study, contributed to the interpretation of the results and wrote the manuscript, S.L. carried the calculations, the simulations, produced the figures and contributed to the interpretation of the results, F.L. helped with the simulations

\noindent{\bf Competing  financial interests:}  The authors declare no competing  financial interests.

\section*{methods}\label{APP}

We first observe that in the light of \eqs (\ref{NMV}) and (\ref{generalRNM})
\begin{eqnarray}
{\int_0^\infty\!{\rm d}t\,\partial_t|\alpha(t)|^4}&=&\mathcal N_V+\int_{\partial_t{|\alpha(t)|}<0}\!\!\!{\rm d}t\,\partial_t|\alpha(t)|^4\nonumber\\
&=&\mathcal N_V-\left|\int_{\partial_t{|\alpha(t)|}<0}\!\!\!{\rm d}t\,\partial_t|\alpha(t)|^4\right|\label{int}\,.\,\,\,
\end{eqnarray} 
Moreover, for an amplitude-damping channel [\cf\eq(\ref{TLSmap})], $\alpha(\infty)=0$ while of course $\alpha(0)=1$. Hence, the leftmost-hand side of \eq(\ref{int}) equals -1 and thus
$$\left|\int_{\partial_t{|\alpha(t)|}<0}\!\!\!{\rm d}t\,\partial_t|\alpha(t)|^4\right|=\mathcal N_V+1\,,$$
which shows \eq(\ref{NV+1}). The calculation of the rescaled non-Markovianity measure $\mathcal N$ thus reduces to that of the non-rescaled measure $\mathcal N_V$. Based on \eq(\ref{NMV}), it is easy to see that this can be expressed as
$$\mathcal N_V=\sum_M |\alpha(t_M)|^4-\sum_m |\alpha(t_m)|^4\,,$$
where $\{t_M\}$ ($\{t_m\}$) are the local maxima (minima) points of the time function $|\alpha(t)|^4$ (volume of accessible states) with
$\alpha(t)$ in our case given by \eq(\ref{alphaEFF}).

For $\omega_\ell=\omega_a$, based on \eq(\ref{alphaEFF}) we see that regardless of $r=\gamma/g_\ell$ all the local minima are zero, i.e., $|\alpha(t_m)|^4=0$. For $0{<}r{<}2\sqrt{2}$,
the local maxima occur at times
\begin{equation}
t_M=\frac{4}{\Delta }\left[\tan^{{-}1}\!\left(\frac{2 \Delta  r}{r^2{-}\Delta ^2}\right){+}\pi  M\right]\;\; \,\,\,\,(M{=}1,2,..)\,.
\label{maxima}\end{equation}
For $2\sqrt{2}{<}r{<}4$, a further maximum occurs at $t=t_0$ with $t_0$ given by \eq(\ref{maxima}) for $M=0$. Finally,
for $r{>}4$ function $|\alpha(t)|^4$ exhibits a single local maximum at $t=t_0$.

\end{document}